\documentclass[12pt]{article}
\usepackage{hyperref}
\usepackage[english]{babel}
\usepackage{subfigure}
\usepackage{amsmath}
\usepackage{latexsym}
\usepackage{amssymb}
\usepackage[all]{xy}
\usepackage[dvips]{graphicx}
\usepackage{epsfig}

\numberwithin{equation}{section} \numberwithin{table}{section}
\numberwithin{figure}{section}

\begin{document}



\newcommand{\tcb}{\textcolor{blue}}
\newcommand{\tcr}{\textcolor{red}}
\newcommand{\tcg}{\textcolor{green}}


\def\be{\begin{equation}}
\def\ee{\end{equation}}
\def\ba{\begin{array}}
\def\ea{\end{array}}
\def\bea{\begin{eqnarray}}
\def\eea{\end{eqnarray}}
\def\nn{\nonumber\\}


\def\ct{\cite}
\def\la{\label}
\def\eq#1{Eq. (\ref{#1})}


\def\a{\alpha}
\def\b{\beta}
\def\g{\gamma}
\def\G{\Gamma}
\def\d{\delta}
\def\D{\Delta}
\def\ep{\epsilon}
\def\e{\eta}
\def\ph{\phi}
\def\Ph{\Phi}
\def\ps{\psi}
\def\Ps{\Psi}
\def\k{\kappa}
\def\l{\lambda}
\def\L{\Lambda}
\def\m{\mu}
\def\n{\nu}
\def\th{\theta}
\def\Th{\Theta}
\def\r{\rho}
\def\s{\sigma}
\def\S{\Sigma}
\def\ta{\tau}
\def\o{\omega}
\def\O{\Omega}
\def\pr{\prime}


\def\half{\frac{1}{2}}

\def\goto{\rightarrow}

\def\na{\nabla}
\def\grad{\nabla}
\def\curl{\nabla\times}
\def\div{\nabla\cdot}
\def\pa{\partial}

\def\ln{\left.}
\def\rn{\right.}
\def\rb{\right]}
\def\rc{\right\}}
\def\rs{\right)}
\def\bra{\left\langle}
\def\ket{\right\rangle}
\def\lb{\left[}
\def\lc{\left\{}
\def\ls{\left(}
\def\lp{\left.}
\def\rp{\right.}

\begin{titlepage}
  \begin{flushright}
  {\small CQUeST-2010-0369}
  \end{flushright}

  \begin{center}

    \vspace{20mm}

    {\LARGE \bf Strange Metallic Behavior in Anisotropic Background}

    \vspace{10mm}

   Bum-Hoon Lee$^{\ast\dag}$, Da-Wei Pang$^{\dag}$ and Chanyong Park$^{\dag}$

    \vspace{5mm}
    {\small \sl $\ast$ Department of Physics, Sogang University}\\
    {\small \sl $\dag$ Center for Quantum Spacetime, Sogang University}\\
    {\small \sl Seoul 121-742, Korea\\}
    {\small \tt bhl@sogang.ac.kr, pangdw@sogang.ac.kr, cyong21@sogang.ac.kr}
    \vspace{10mm}

  \end{center}

\begin{abstract}
\baselineskip=18pt We continue our analysis on conductivity in the
anisotropic background by employing the D-brane probe technique,
where the D-branes play the role of charge carriers. The DC and AC
conductivity for massless charge carriers are obtained analytically,
while interesting curves for the AC conductivity are also plotted.
For massive charge carriers, we calculate the DC and AC
conductivities in the dilute limit and we fix the parameters in the
Einstein-Maxwell-dilaton theory so that the background exhibits the
same scaling behaviors as those for real-world strange metals. The
DC conductivity at finite density is also computed.

\end{abstract}
\setcounter{page}{0}
\end{titlepage}

\pagestyle{plain} \baselineskip=19pt

\tableofcontents

\section{Introduction}
The AdS/CFT correspondence~\cite{Maldacena:1997re, Aharony:1999ti}
provides us with powerful techniques for analyzing dynamics of
strongly coupled quantum field theories. Properties of certain
strongly coupled quantum field theories, such as QCD, can be
inferred from the dual gravity side. Therefore the AdS/CFT
correspondence may shed light on understanding physical systems in
the real world via holography. Recently, inspired by condensed
matter physics, the investigations on the applications of AdS/CFT to
condensed matter systems have accelerated enormously. Some excellent
reviews are given by~\cite{Hartnoll:2009sz}.

By now, there are mainly two complementary approaches for studying
this AdS/Condensed Matter Theory(CMT) correspondence. One can be
seen as the bottom-up approach, that is, we investigate some
phenomenological models of gravity which can exhibit similar
interesting behaviors of certain condensed matter systems. One
specific example is the construction of holographic
superconductor~\cite{Gubser:2008px, Hartnoll:2008vx, Gubser:2008wv,
Hartnoll:2008kx, Chen:2010mk, Herzog:2010vz}. The other one is the
top-down approach, which means that we consider some background in
string/M theory and study the corresponding properties using string
theory techniques. One can also investigate holographic
superconductors in the context of string/M theory, for examples see
~\cite{Ammon:2008fc, Ammon:2009fe, Gubser:2009qm, Gauntlett:2009dn,
Kaminski:2010zu}.

Recently a holographic model building approach to `strange metallic'
phenomenology was proposed in~\cite{Hartnoll:2009ns}. They
considered a bulk gravitational background which was dual to a
neutral Lifshitz-invariant quantum critical theory, where the gapped
probe charge carriers were described by D-branes. The non-Fermi
liquid scalings, such as linear resistivity, observed in strange
metal regimes can be realized by choosing the dynamical critical
exponent $z$ appropriately. They also outlined three distinct string
theory realizations of Lifshitz geometries. In this paper we will
study conductivities in charged dilaton black hole background with
anisotropic scaling symmetry by employing the probe D-brane method.

Charged dilaton black holes~\cite{Gibbons:1987ps, Preskill:1991tb,
Garfinkle:1990qj, Holzhey:1991bx} for the Einstein-Maxwell theory
with exponential dilaton coupling $\mathrm{e}^{2 \alpha \phi} F^2$
possess several interesting properties. First, it has been found
that for any values of $\alpha \neq 0$, the entropy vanishes in the
extremal limit, which means that the thermodynamic description might
break down. Second, when the coupling $\alpha$ takes certain
specific value, the corresponding black hole solution can be
embedded into string theory. The peculiar features of such charged
dilaton black holes suggest that their AdS generalizations may
provide interesting holographic descriptions of condensed matter
systems.

Recently, holography of charged dilaton black holes in AdS$_4$ with
planar symmetry was extensively investigated
in~\cite{Goldstein:2009cv}.
The near horizon geometry was Lifshitz-like~\cite{Kachru:2008yh,
Taylor:2008tg}, with a dynamical exponent $z$ determined by the
dilaton coupling. The global solution was constructed via numerical
methods and the attractor behavior was also discussed. The authors
also examined the thermodynamics of near extremal black holes and
computed the AC conductivity in zero-temperature background. For
related work on charged dilaton black holes see~\cite{Gubser:2009qt,
Gauntlett:2009bh, Cadoni:2009xm, Chen:2010kn, Charmousis:2010zz}.

Here we consider charged dilaton black holes in
Einstein-Maxwell-dilaton theory where the scalar potential takes a
Liouville form. Charged dilaton black holes with a Liouville
potential were studied, e.g. in~\cite{Cai:1996eg}. In a previous
paper~\cite{Lee:2010xx}, we calculated conductivities in such
charged dilaton black hole backgrounds via conventional methods. In
this paper we will continue our investigations on conductivities in
such backgrounds by studying the dynamics of probe D-branes,
following~\cite{Hartnoll:2009ns}. One key feature of our
investigations is that we incorporate a non-trivial dilaton in the
DBI action. As was pointed out in~\cite{Hartnoll:2009ns}, such a
dilaton field might be helpful in the holographic model building of
strange metals. Indeed we find that when the parameters in the
Einstein-Maxwell-dilaton theory take certain specific values, the
resistivity and conductivity exhibit scaling behaviors of real-world
``strange'' metals.

The rest of the paper is organized as follows: In Section 2 we will
present the charged dilaton black hole solutions with a Liouville
potential. The DC and AC conductivities will be calculated in
Section 3 using the probe brane technique. We consider both the
massless and massive charge carriers and find strange metal-like
behaviors in the above mentioned background. Summary and discussion
will be given in the final section.

\section{Solution with a Liouville potential}
In this section, we review the construction of a background
space-time containing an anisotropic scaling at the boundary, which
was studied in our previous work~\cite{Lee:2010xx}. Consider the
following action
\begin{equation}    \la{orgact}
S=\int
d^{4}x\sqrt{-g}[R-2(\nabla\phi)^2-e^{2\alpha\phi}F_{\mu\nu}F^{\mu\nu}-V(\ph)],
\end{equation}
where $\ph$ and $V(\ph)$ represent a dilaton field and its
potential. Equations of motion for metric $g_{\m\n}$, dilaton field
and U(1) gauge field are
\begin{eqnarray}
R_{\mu\nu}-\frac{1}{2}Rg_{\mu\nu}+\frac{1}{2}g_{\mu\nu}V(\phi) &=&
2\partial_{\mu}\phi
\partial_{\nu}\phi-g_{\mu\nu}(\nabla\phi)^{2}+2e^{2\alpha\phi}F_{\mu\lambda}{F_{\nu}}^{\lambda}
-\frac{1}{2}g_{\mu\nu}e^{2\alpha\phi}F^{2}, \\
\partial_{\mu}(\sqrt{-g}\partial^{\mu}\phi) &=& \frac{1}{4}\sqrt{-g}\frac{\partial
V(\phi)}{\partial\phi}+\frac{\alpha}{2}\sqrt{-g}e^{2\alpha\phi}F^{2},~~~ \\
0 &=& \partial_{\mu}(\sqrt{-g}e^{2\alpha\phi}F^{\mu\nu})
\la{eqgauge}.
\end{eqnarray}
Now, we choose a Liouville-type potential as a dilaton potential
\be\label{2eq5} V(\phi)=2\Lambda e^{ \eta\phi} . \ee For $\eta=0$,
the Liouville potential reduces to a cosmological constant, which
was studied in Ref. \cite{Goldstein:2009cv}. To solve equations of
motion, we use the following scaling ansatz which corresponds to a
zero temperature solution
\begin{equation}
ds^{2}=-a(r)^{2}dt^{2}+\frac{dr^{2}}{a(r)^{2}}+b(r)^{2}(dx^{2}+dy^{2})
,
\end{equation}
with
\begin{equation}
a(r)=t_{0}r^{a_{1}},~~~b(r)=b_{0}r^{b_{1}},~~~\phi(r)=-k_{0}\log r.
\end{equation}
If we turn on a time-component of the gauge field $A_t$ only, from
the above metric ansatz the electric flux satisfying \eq{eqgauge}
becomes
\begin{equation}
F_{tr}=\frac{q}{b(r)^{2}}e^{-2\alpha\phi}.
\end{equation}
The rest of equations of motion are satisfied when various
parameters appeared in the above are given by \bea    \la{param} &&
a_{1}=1 -
\frac{k_{0}}{2}\eta,~~~b_{1}=\frac{(2\alpha+\eta)^{2}}{(2\alpha+\eta)^{2}+16},
~~~k_{0}=\frac{4(2\alpha+\eta)}{(2\alpha+\eta)^{2}+16}, ~~~b_{0}=1 ,
\nn &&
t_{0}^{2}=\frac{-2\Lambda}{(a_{1}+b_{1})(2a_{1}+2b_{1}-1)},~~~
q^{2}=-   \frac{2k_{0} \L}{\a \ls a_{1}+b_{1} \rs} +\frac{\eta}{2}
\frac{\L}{\a} , \eea where we consider a negative $\L$ only. Note
that the above solution is an exact solution to the equations of
motion containing three parameters $\a$, $\eta$ and $\L$ and the
parameter $b_1$ is always smaller than $1$. Especially, for $\eta=0$
and $\L = -3/L^{2}$ this solution reduces to the one in Ref.
\cite{Goldstein:2009cv}, as previously mentioned. We will set $L=1$
in the following for simplicity and we will restore the factor of
$L$ in the conductivities by dimensional analysis. For $2\a = -
\eta$, the above solution becomes $AdS_2 \times R^2$. If we take a
limit, $\a \to \infty$, and at the same time set $q=\eta=0$, we can
obtain $AdS_4$ geometry. When $\eta$ is proportional to $\a$ like
$\eta = - c \a$, the metric in the limit, $\a \to \infty$, is
reduced to \be ds^2 = - t_0^2 r^{2 z} dt^2 + \frac{dr^2}{t_0^2
r^{2z}} + r^2 \ls dx^2 + dy^2 \rs , \ee with \be z = \frac{2+c}{2-c}
. \ee The power $z$ is given by $2$ for $c=2/3$ and $3$ for $c=1$,
etc.

The previous solution can be easily extended to a finite temperature
case corresponding to black hole. With the same parameters in
\eq{param}, the black hole solution becomes \be \la{anisomet}
ds^{2}=-a(r)^{2}f(r)dt^{2}+\frac{dr^{2}}{a(r)^{2}f(r)}+b(r)^{2}(dx^{2}+dy^{2}),
\ee where \be
f(r)=1-\frac{r^{2a_{1}+2b_{1}-1}_{+}}{r^{2a_{1}+2b_{1}-1}} .
\end{equation}
Notice that since the above black hole factor does not include the
black hole charge this solution does not correspond to the
Reissner-Nordstrom but Schwarzschild black hole. The Hawking
temperature of this black hole is given by
\begin{equation}
T=\frac{1}{4\pi}(2a_{1}+2b_{1}-1) t_{0}^{2}r_{+}^{2a_{1}-1}.
\end{equation}where
$r_+$ means the position of the black hole horizon. The
conductivities in such backgrounds, both at zero temperature and
finite temperature, were investigated in our previous work using
conventional techniques~\cite{Lee:2010xx}.
Following~\cite{Hartnoll:2009ns}, in the subsequent sections we will
discuss the conductivities induced by the massless and massive
charge carriers, which are represented by the probe D$q$-branes in
the charged dilaton black hole background.

\section{Massless charge carriers}
As was advocated in~\cite{Hartnoll:2009ns}, once we started to
investigate mechanisms for strange metal behaviors by holographic
technology, we should involve a sector of (generally massive) charge
carriers carrying nonzero charge density $J^{t}$, interacting with
themselves and with a larger set of neutral quantum critical degrees
of freedom. When interpreted in the context of D-branes, the sector
of charge carriers were modeled by probe ``flavor branes'' in the
spirit of~\cite{Karch:2002sh}. The conductivity of such charge
carriers was obtained in~\cite{Karch:2007pd} in an elegant way. The
main point is that both the numerator and the denominator in the
square root of the on-shell DBI action for the probe brane change
sign between the horizon $v=v_{+}$ and the boundary $v=0$, therefore
to ensure the reality of the action the sign change must take at the
same radial position $0<v_{\ast}<v_{+}$. Then the conductivity can
be read off by the Ohm's law after solving the equations coming from
the numerator and the denominator respectively. The Hall
conductivity can be obtained in a similar
way~\cite{O'Bannon:2007in}. We focus on the case of massless charge
carriers in this section, while the massive case will be considered
in the next section.

\subsection{DC conductivity}
After taking the following coordinate transformation
\begin{equation}
v=\frac{1}{r},~~~v_{+}=\frac{1}{r_{+}},
\end{equation}
the metric~(\ref{anisomet}) becomes
\begin{equation}
ds^{2}=-\frac{a_{0}^{2}}{v^{2a_{1}}}f(v)dt^{2}+\frac{v^{2a_{1}-4}}{a_{0}^{2}f(v)}dv^{2}
+\frac{1}{v^{2b_{1}}}(dx^{2}+dy^{2}),
\end{equation}
where
\begin{equation}
f(v)=1-\frac{v^{\delta}}{v^{\delta}_{+}},~~~\delta=2a_{1}+2b_{1}-1.
\end{equation}
The temperature of the black hole can be rephrased as
\begin{equation}
\label{3eq4} T=\frac{1}{4\pi}\delta a_{0}^{2}v_{+}^{1-2a_{1}}.
\end{equation}

Next we consider the probe D$q$-brane
\begin{equation}
S_{q}=-T_{q}\int d\tau d^{q}\sigma e^{-\phi}\sqrt{-{\rm
det}(g_{ab}+2\pi\alpha^{\prime}F_{ab})},
\end{equation}
where $T_{q}=(g_{s}(2\pi)^{q}l_{s}^{q+1})^{-1}$ is the D$q$-brane
tension. Notice that the Wess-Zumino terms are neglected here. The
embedding of the probe brane can be described as follows
\begin{equation}
\tau=t,~~~\sigma^{1}=x,~~~\sigma^{2}=y,~~~\sigma^{3}=v,~~~\{\sigma^{4},\cdots,\sigma^{q}\}=\Sigma.
\end{equation}
Therefore the DBI action can be rewritten as
\begin{equation}
\label{3eq7} S_{q}=-\tau_{\rm eff}\int
dtdvd^{2}xv^{-k_{0}}\sqrt{-{\rm
det}(g_{ab}+2\pi\alpha^{\prime}F_{ab})},
\end{equation}
where $\tau_{\rm eff}=T_{q}{\rm Vol}(\Sigma)$, ${\rm Vol}(\Sigma)$
denoting the volume of the compact manifold. Note that here the
dilaton has a non-trivial dependence on the radial coordinate $v$,
$e^{-\phi}=v^{-k_{0}}$. As was emphasized in~\cite{Hartnoll:2009ns},
incorporating a non-trivial dilaton might lead to a more realistic
holographic model of strange metals and we will see that this is
indeed the case.

We take the following ansatz for the worldvolume $U(1)$ gauge field
\begin{equation}
A=\Phi(v)dt+(-Et+h(v))dx.
\end{equation}
Then the DBI action~(\ref{3eq7}) becomes
\begin{equation}
S_{q}=-\tau_{\rm eff}\int
dtd^{2}xdvv^{-k_{0}}\sqrt{g_{xx}}\sqrt{-g_{tt}g_{xx}g_{vv}-(2\pi
\alpha^{\prime})^{2}(g_{vv}E^{2}+g_{xx}\Phi^{\prime2}+g_{tt}h^{\prime2})}.
\end{equation}
It can be seen that the above action depends only on $\Phi^{\prime}$
and $h^{\prime}$, which leads to two conserved quantities
\begin{equation}
C=\frac{-g_{xx}^{3/2}\Phi^{\prime}v^{-k_{0}}}{\sqrt{-g_{tt}g_{xx}g_{vv}-(2\pi
\alpha^{\prime})^{2}(g_{vv}E^{2}+g_{xx}\Phi^{\prime2}+g_{tt}h^{\prime2})}},
\end{equation}
\begin{equation}
H=\frac{-g_{tt}g_{xx}^{1/2}h^{\prime}v^{-k_{0}}}{\sqrt{-g_{tt}g_{xx}g_{vv}-(2\pi
\alpha^{\prime})^{2}(g_{vv}E^{2}+g_{xx}\Phi^{\prime2}+g_{tt}h^{\prime2})}}.
\end{equation}
One can see that here these quantities satisfy the relation
$g_{tt}h^{\prime}C=g_{xx}\Phi^{\prime}H$.

Following~\cite{Karch:2007pd}, we can solve for $\Phi^{\prime}$ and
$ h^{\prime}$ from the above two equations and substitute the
solutions back into the action. The on-shell DBI action turns out to
be
\begin{equation}
S_{q}=-\tau_{\rm eff}\int dtd^{2}xdv
v^{-2k_{0}}g_{xx}^{3/2}\sqrt{-g_{tt}g_{vv}}\sqrt{\frac{g_{tt}g_{xx}+(2\pi\alpha^{\prime})^{2}E^{2}}
{(2\pi\alpha^{\prime})^{2}(C^{2}g_{tt}+H^{2}g_{xx})+g_{xx}^{2}g_{tt}v^{-2k_{0}}}}.
\end{equation}
As pointed out in~\cite{Karch:2007pd}, both the numerator and the
denominator in the square root change sign between the boundary
$v=0$ and the horizon $v=v_{+}$. To make sure that the action is
real, the numerator and the denominator should change sign at the
same radial position $0<v_{\ast}<v_{+}$, which requires
\begin{equation}
\label{3eq13}
g_{tt}g_{xx}+(2\pi\alpha^{\prime})^{2}E^{2}\big|_{v=v_{\ast}}=0,
\end{equation}
\begin{equation}
\label{3eq14}
(2\pi\alpha^{\prime})^{2}(C^{2}g_{tt}+H^{2}g_{xx})+g_{xx}^{2}
g_{tt}v^{-2k_{0}}\big|_{v=v_{\ast}}=0.
\end{equation}
The first equation gives
\begin{equation}
a^{2}_{0}f(v_{\ast})=(2\pi\alpha^{\prime})^{2}E^{2}v^{2a_{1}+2b_{1}}_{\ast}.
\end{equation}

Following~\cite{Hartnoll:2009ns}, we identify the constants of
motion with the currents as follows
\begin{equation}
J^{x}=(2\pi\alpha^{\prime})^{2}\tau_{\rm
eff}H,~~~J^{t}=(2\pi\alpha^{\prime})^{2}\tau_{\rm eff}C.
\end{equation}
Therefore~(\ref{3eq14}) can be rewritten in the desired form
\begin{equation}
J^{x}=E\sqrt{(2\pi\alpha^{\prime})^{4}\tau_{\rm
eff}^{2}v^{-2k_{0}}_{\ast}+(2\pi\alpha^{\prime})^{2}(J^{t})^{2}v^{4b_{1}}_{\ast}}.
\end{equation}
Finally, according to Ohm's law $J^{x}=\sigma E$, the conductivity
is given by
\begin{equation}
\label{3eq18} \sigma=\sqrt{(2\pi\alpha^{\prime})^{4}\tau_{\rm
eff}^{2}v^{-2k_{0}}_{\ast}+(\frac{2\pi\alpha^{\prime}}{L^{2}})^{2}(J^{t})^{2}v^{4b_{1}}_{\ast}}.
\end{equation}
There exist two terms in the square root. One may
interpret the first term as arising from thermally produced pairs of
charge carriers, though here it has some non-trivial dependence on
$v_{\ast}$. It is expected that such a term should be suppressed
when the charge carriers have large mass. Then the surviving term
gives
\begin{equation}
\sigma=\frac{2\pi\alpha^{\prime}}{L^{2}}J^{t}v^{2b_{1}}_{\ast}.
\end{equation}
By combining~(\ref{3eq4}), one can obtain the power-law for the DC
resistivity,
\begin{equation}
\rho\sim\frac{T^{\lambda}}{J^{t}},~~~\lambda=\frac{2b_{1}}{2a_{1}-1},
\end{equation}
where we take the limit $E\ll1$ so that $v_{\ast}\approx v_{+}$.
Note that when the parameter $\eta$ in the Liouville
potential~(\ref{2eq5}) is zero, the background reduces to be a
Lifshitz-like solution at finite temperature with $a_{1}=1$ and
$b_{1}=z^{-1}$. Thus we have $\rho\sim T^{2/z}/J^{t}$, which agrees
with the result obtained in~\cite{Hartnoll:2009ns}.
\subsection{DC Hall conductivity}
We can also calculate the conductivity tensor
$$J^{i}=\sigma^{ij}E_{j}$$ by generalizing the techniques presented
in~\cite{Karch:2007pd}. Hall conductivities for general D$p$-D$q$
systems were obtained in~\cite{O'Bannon:2007in}. Similarly here we
can take the following ansatz for the worldvolume $U(1)$ gauge
fields
\begin{equation}
A_{t}=\Phi(v),~~~A_{x}(v,t)=-Et+f_{x}(v),~~~A_{y}(v,x)=Bx+f_{y}(v).
\end{equation}
The DBI action of the probe brane is still given by
\begin{equation}
S_{q}=-\tau_{\rm eff}\int dtdvd^{2}xv^{-k_{0}}\sqrt{-{\rm
det}(g_{ab}+2\pi\alpha^{\prime}F_{ab})},
\end{equation}
where
\begin{eqnarray}
-{\rm
det}(g_{ab}+2\pi\alpha^{\prime}F_{ab})&=&-g_{tt}g_{vv}g_{xx}^{2}
-(2\pi\alpha^{\prime})^{2}g_{tt}g_{vv}B^{2}-(2\pi\alpha^{\prime})^{2}g_{tt}g_{xx}f_{x}^{\prime2}
\nonumber\\ &
&-(2\pi\alpha^{\prime})^{2}g_{tt}g_{xx}f_{y}^{\prime2}-(2\pi\alpha^{\prime})^{2}g_{vv}g_{xx}E^{2}
-(2\pi\alpha^{\prime})^{2}g_{xx}^{2}\Phi^{\prime2}\nonumber\\
&
&-(2\pi\alpha^{\prime})^{4}\Phi^{\prime2}B^{2}-(2\pi\alpha^{\prime})^{4}f^{\prime2}_{y}E^{2}
+2(2\pi\alpha^{\prime})^{4}EB\Phi^{\prime}f_{y}^{\prime}.
\end{eqnarray}

Now we have three conserved quantities
\begin{equation}
C=\frac{-g_{xx}^{2}\Phi^{\prime}-(2\pi\alpha^{\prime})^{2}B^{2}\Phi^{\prime}
+(2\pi\alpha^{\prime})^{2}EBf_{y}^{\prime}}{\sqrt{-{\rm
det}(g_{ab}+2\pi\alpha^{\prime}F_{ab})}}v^{-k_{0}},
\end{equation}
\begin{equation}
H=\frac{-g_{tt}g_{xx}f^{\prime}_{x}}{\sqrt{-{\rm
det}(g_{ab}+2\pi\alpha^{\prime}F_{ab})}}v^{-k_{0}},
\end{equation}
\begin{equation}
M=\frac{-g_{tt}g_{xx}f_{y}^{\prime}-(2\pi\alpha^{\prime})^{2}E^{2}f_{y}^{\prime}
+(2\pi\alpha^{\prime})^{2}EB\Phi^{\prime}}{\sqrt{-{\rm
det}(g_{ab}+2\pi\alpha^{\prime}F_{ab})}}v^{-k_{0}},
\end{equation}
To calculate the on-shell DBI action, we should first express
$f^{\prime}_{x,y}$ in terms of $\Phi^{\prime}$. Then we solve the
equation of motion for $\Phi^{\prime}$ and substitute them back into
$-{\rm det}(g_{ab}+2\pi\alpha^{\prime}F_{ab})$. The on-shell DBI
action takes the following simple form
\begin{equation}
S_{q}=\tau_{\rm eff}\int
dtd^{2}xdvv^{-k_{0}}\sqrt{-g_{tt}g_{vv}}g_{xx}\frac{\xi}{\sqrt{\xi\chi-a^{2}}},
\end{equation}
where
\begin{equation}
\xi=-[(2\pi\alpha^{\prime})^{2}E^{2}g_{xx}+(2\pi\alpha^{\prime})^{2}B^{2}g_{tt}+g_{tt}g^{2}_{xx}],
\end{equation}
\begin{equation}
\chi=-g_{tt}g_{xx}^{2}-(2\pi\alpha^{\prime})^{2}[g_{tt}C^{2}+g_{xx}(H^{2}+M^{2})]v^{2k_{0}},
\end{equation}
\begin{equation}
a=(2\pi\alpha^{\prime})^{2}(MEg_{xx}-BCg_{tt})v^{k_{0}},
\end{equation}

As argued in~\cite{O'Bannon:2007in}, the only consistent possibility
for preserving the reality of the on-shell DBI action is to require
$\xi, \chi$ and $a$ to become zero at the same radial position
$v_{\ast}$. Therefore by solving the equations $\xi,\chi,a=0$ and
identifying
\begin{equation}
J^{x}=(2\pi\alpha^{\prime})^{2}\tau_{\rm
eff}H,~~~J^{t}=(2\pi\alpha^{\prime})^{2}\tau_{\rm eff}C,~~~ J^{y}
=(2\pi\alpha^{\prime})^{2}\tau_{\rm eff}M,
\end{equation}
one can finally obtain
\begin{equation}
\sigma^{xx}=\frac{v^{-k_{0}}_{\ast}}{1+(\frac{2\pi\alpha^{\prime}}{L^{2}})^{2}B^{2}v^{4b_{1}}_{\ast}}
\sqrt{(2\pi\alpha^{\prime})^{4}\tau_{\rm
eff}^{2}(1+(\frac{2\pi\alpha^{\prime}}{L^{2}})^{2}B^{2}v^{4b_{1}}_{\ast})+(
\frac{2\pi\alpha^{\prime}}{L^{2}})^{2}(J^{t})^{2}
v^{4b_{1}+2k_{0}}_{\ast}},
\end{equation}
and
\begin{equation}
\sigma^{xy}=\frac{(2\pi\alpha^{\prime})^{2}BJ^{t}v^{4b_{1}}_{\ast}}
{L^{4}+(2\pi\alpha^{\prime})^{2}B^{2}v^{4b_{1}}_{\ast}}.
\end{equation}
Here are some remarks on the results for the conductivity tensor.
\begin{itemize}
\item When both $B$ and $E$ are small, the Hall conductivity
becomes $\sigma^{xy}\sim T^{4b_{1}/(1-2a_{1})}$. Once we take
$\eta=0$ in the Liouville potential, we have $a_{1}=1$ and therefore
$\sigma^{xy}\sim T^{-4b_{1}}$. Note that $b_{1}=1/z$, so we recover
the result obtained in~\cite{Hartnoll:2009ns} $\sigma^{xy}\sim
T^{-4/z}$.
\item The expression for $\sigma^{xx}$ reduces to the one obtained
in previous subsection when $B=0$. Furthermore, when the second term
in the square root dominates, and $B$ is small, we reproduce the
result $\sigma^{xx}\sim T^{-\lambda}$ where
$\lambda=\frac{2b_{1}}{2a_{1}-1}$.
\item One interesting quantity for studying the strange metals is
the ratio $\sigma^{xx}/\sigma^{xy}$. When the first term in the
square root of $\sigma^{xx}$ is subdominant, one can easily obtain
the following result
\begin{equation}
\frac{\sigma^{xx}}{\sigma^{xy}}\sim
v^{-2b_{1}}_{\ast}=T^{\frac{-2b_{1}}{1-2a_{1}}}.
\end{equation}
In the $\eta=0$ limit, $a_{1}=1, b_{1}=1/z$, we have
$$\frac{\sigma^{xx}}{\sigma^{xy}}\sim
v^{-2/z}_{\ast}=T^{2/z}.$$
\item The strange metals exhibit the following anomalous behaviors:
$\sigma^{xx}\sim T^{-1}, \sigma^{xx}/\sigma^{xy}\sim T^{2}$. In
contrast, $\sigma^{xx}/\sigma^{xy}\sim(\sigma^{xx})^{-1}$ in Drude
theory. Since $\sigma^{xx}\sim T^{-2/z}$ in the limit of $\eta=0$,
our result can mimic Drude theory in this limit, which agrees with
the result in~\cite{Hartnoll:2009ns}.
\end{itemize}
\subsection{AC conductivity}
The AC conductivity will be calculated in this subsection. Rather
than working out the full nonlinear dependence on the electric
field, here we will expand in small fluctuations of the background
gauge field. The background gauge field can be obtained by setting
$E=H=0$,
\begin{equation}
\Phi^{\prime}=\frac{C}{\gamma}v^{2b_{1}+k_{0}-2},~~~\gamma\equiv\sqrt{1
+(2\pi\alpha^{\prime})^{2}C^{2}v^{2k_{0}+4b_{1}}},
\end{equation}
Consider the fluctuations of the probe gauge fields as the following
form
\begin{equation}
\delta A=(A_{t}(v)dt+A_{x}(v)dx+A_{y}(v)dy)e^{-i(\omega t-kx)},
\end{equation}
The DBI action can be expanded as
\begin{eqnarray}
S_{q}&=&-\tau_{\rm eff}\int
dtd^{2}xdvv^{-k_{0}}\gamma^{-1}\sqrt{-g_{tt}g_{vv}}g_{xx}[1+\frac{1}{2}\gamma
(2\pi\alpha^{\prime})^{2}\frac{F^{2}_{xy}}{g^{2}_{xx}}\nonumber\\& &
+\frac{1}{2}\gamma^{2}
(2\pi\alpha^{\prime})^{2}\frac{F^{2}_{iv}}{g_{vv}g_{xx}}-\frac{1}{2}\gamma^{2}
(2\pi\alpha^{\prime})^{2}\frac{F^{2}_{ti}}{g_{tt}g_{xx}}-\frac{1}{2}\gamma^{3}
(2\pi\alpha^{\prime})^{2}\frac{F^{2}_{tv}}{g_{tt}g_{vv}}],
\end{eqnarray}
Let us focus on the quadratic terms of $F_{\mu\nu}$
\begin{eqnarray}
S_{F}&=&-\frac{\tau_{\rm eff}}{2}(2\pi\alpha^{\prime})^{2}\int
dtd^{2}xdvv^{-k_{0}}\gamma[\frac{\sqrt{-g_{tt}g_{vv}}}{g_{xx}\gamma}
F^{2}_{xy}\nonumber\\& &+\sqrt{\frac{-g_{tt}}{g_{vv}}}F^{2}_{iv}
-\sqrt{\frac{-g_{vv}}{g_{tt}}}F^{2}_{ti}-\frac{\gamma
g_{xx}}{\sqrt{-g_{tt}g_{vv}}}F^{2}_{tv}],
\end{eqnarray}
from which we can derive the equation of motion for $A_{x}$
\begin{equation}
\partial_{v}(v^{-k_{0}}\sqrt{\frac{-g_{tt}}{g_{vv}}}\gamma A_{x}^{\prime})
=-v^{-k_{0}}\sqrt{\frac{g_{vv}}{-g_{tt}}}\gamma\omega^{2}A_{x},
\end{equation}
It has been pointed out in~\cite{Gubser:2008wz, Horowitz:2009ij}
that the equation of motion for the fluctuation can be converted
into a Schr\"{o}dinger equation. For our case this can be realized
by defining
\begin{equation}    \la{eqs}
A_{x}=(v^{-k_{0}}\gamma)^{-1/2}\Psi,~~~\frac{d}{dv}=\frac{v^{2a_{1}-2}}{a^{2}_{0}f(v)}\frac{d}{ds}.
\end{equation}
Then the equation of motion for $A_x$ becomes a Schr\"{o}dinger form
\begin{equation}    \la{eqps}
-\frac{d^{2}}{ds^{2}}\Psi+U\Psi=\omega^{2}\Psi,
\end{equation}
with the following effective potential
\begin{equation}
U=\frac{1}{2}\frac{1}{\sqrt{v^{-k_{0}}\gamma}}\frac{d}{ds}[\frac{1}{\sqrt{v^{-k_{0}}\gamma}}
\frac{d}{ds}(v^{-k_{0}}\gamma)].
\end{equation}

\begin{figure} \la{fig1}
\begin{center}
\vspace{3cm} \hspace{-0.5cm} \subfigure{
\includegraphics[angle=0,width=0.5\textwidth]{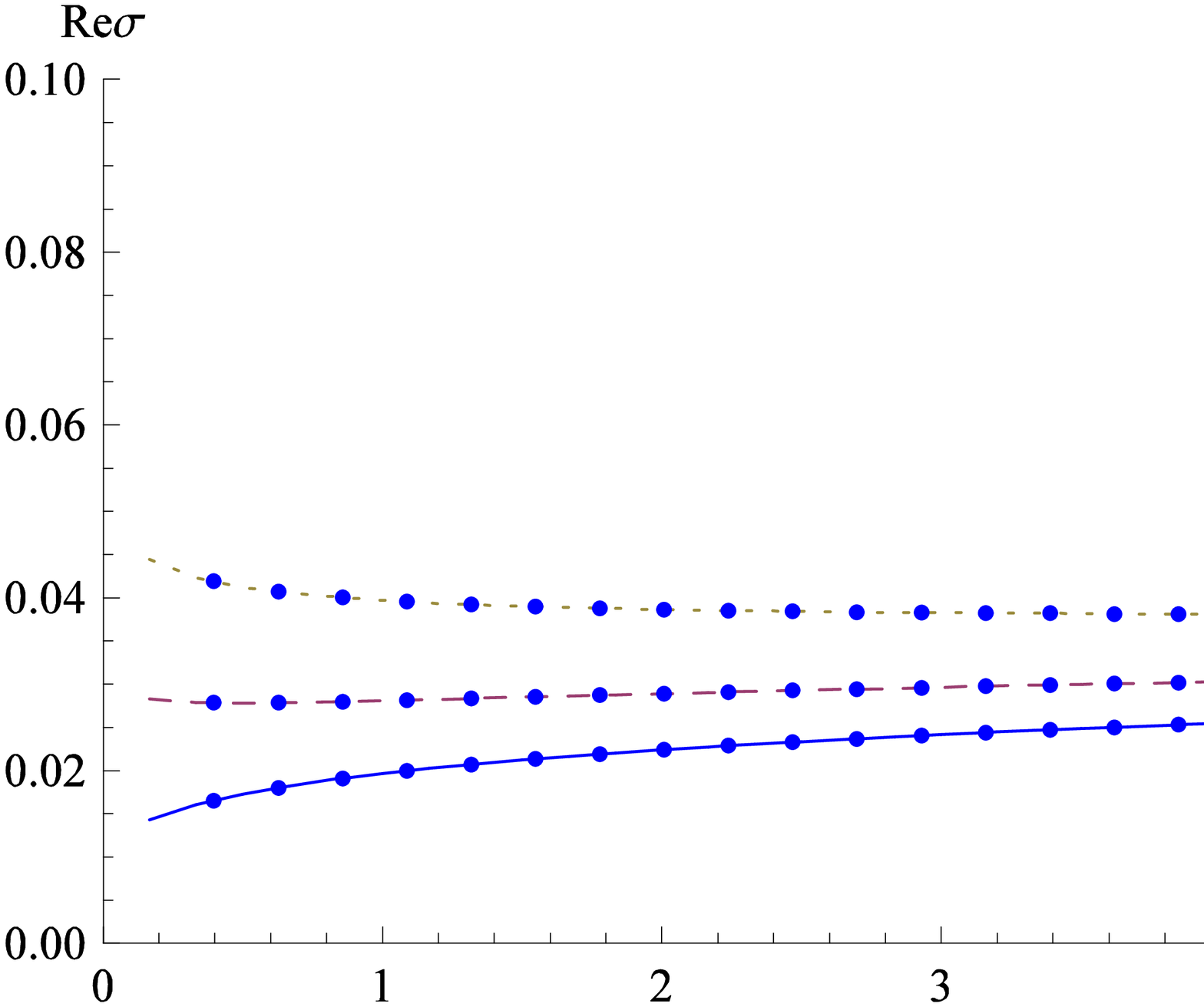}}
\hspace{-0.5cm} \subfigure{
\includegraphics[angle=0,width=0.5\textwidth]{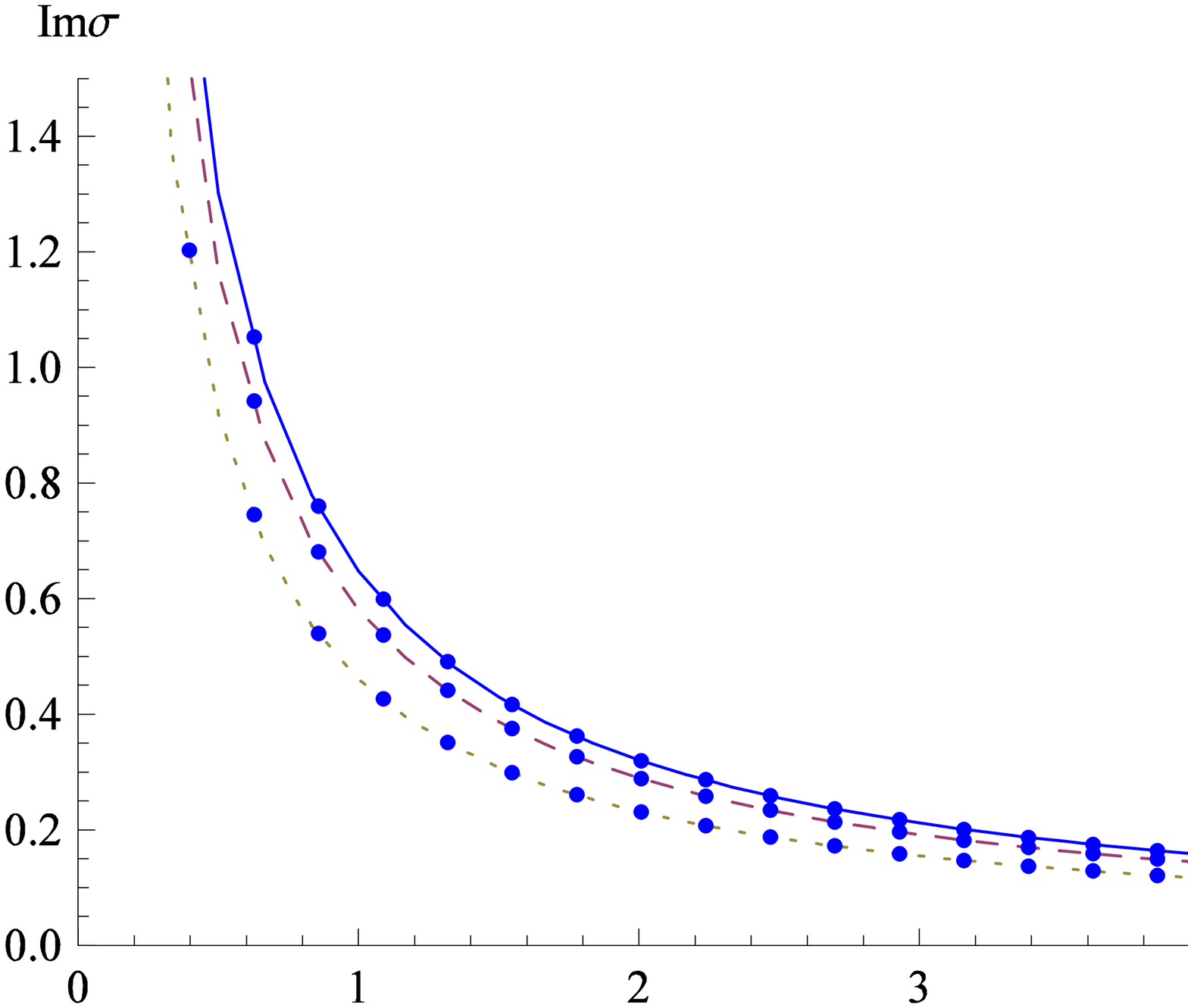}}
\vspace{-2.5cm} \\
\caption{\small The real and imaginary conductivity for
$C=0$(solid), $2$(dashed) and $4$(dotted), where we set $2 \pi
\a'=1$, $\a = - \eta = 1$, $\L = - 3$, $\tau_{\rm eff} = 1$ and $v_+
= 10$.} \label{fig1}
\end{center}
\end{figure}

For numerical calculation, $s$ in \eq{eqs} can be exactly written in terms of $v$
\be
s = \frac{v^{2 a_1 -1}}{a_0^2 (2 a_1 - 1)}  \ {}_2 F_1 \ls \frac{2 a_1 -1}{\d} , 1,
1 + \frac{2 a_1 -1}{\d} , \frac{v^{\d}}{v_+^{\d}} \rs .
\ee
Near the boundary and horizon, $s$ has the following behaviors
\bea
s &\approx& \frac{v^{2 a_1 -1}}{a_0^2 (2 a_1 -1)} \qquad {\rm for} \ v \to 0 , \nn
s &\approx& - \frac{v_+^{2 a_1 -1} \log(v_+ - v)}{a_0^2 (2 a_1 + 2 b_1 -1)} \qquad {\rm for} \ v \to v_+ ,
\eea
so $s=\infty$ or $s=0$ corresponds to the horizon or boundary respectively.
Using the above asymptotic behaviors, we can solve the above Schr\"{o}dinger equation numerically.
At the horizon, the leading term of the effective potential $U$ becomes
\be
U \sim \exp \ls - \frac{a_0^2 (  2 a_1 + 2 b_1 - 1) }{ v_+^{ 2 a_1 -1 }} s \rs .
\ee
So at the horizon the Schr\"{o}dinger equation is simply reduced to
\be
 0 = \frac{d^{2}}{ds^{2}} \Psi  + \omega^{2}\Psi,
\ee whose solution satisfying the incoming boundary condition is
given by \be \la{horsol} \Psi = c e^{i \omega s} , \ee where $c$ is
an integration constant. Near the boundary, in the following
parameter range $a_1 > \half$ and $2 b_1 + k_0 > 0$, $U$  becomes
\be U \approx \frac{k_0 ( 4 a_1 + k_0 - 2) }{4 (2 a_1 -1)^2 s^2} .
\ee So the asymptotic solution of the Schr\"{o}dinger equation near
the boundary is given by \be    \la{asympsol} \Ps = c_1 s^{-
\frac{k_0}{4 a_1 - 2}} + c_2 s^{1 + \frac{k_0}{4 a_1 - 2}} . \ee
From this together with \eq{eqs} the asymptotic solution for $A_x$
becomes \be \la{acsol} A_x = A_0 \ls 1 + \frac{c_2}{c_1} (2 a_1
-1)^{-\frac{2 a_1 -1 +k_0}{2 a_1 -1}} a_0^{-\frac{2 (2 a_1 -1
+k_0)}{2 a_1 -1}} v^{2 a_1 -1 + k_0} \rs . \ee where we set \be c_1
= \frac{A_0}{(2 a_1 -1)^{\frac{k_0}{2( 2 a_1 -1)}} a_0^{\frac{
k_0}{2 a_1 -1}}} . \ee In the above, $A_0$ is the boundary value of
$A_x$, which corresponds to the source for the current operator
$J^x$ and the coefficient of the second term corresponds to the vev
for the current operator. When we introduce a background electric
field $E_x (t) \equiv E_x (\omega) e^{- i \omega t}$, the asymptotic
form of $A_x (\omega)$ can be rewritten as \ct{Hartnoll:2009ns} \be
\la{poleq} A_x (\omega) = \frac{E_x (\omega)}{i \omega} + \frac{J_x
(\omega)}{\tau_{\rm eff} (2 \pi \a')^2} \  v^{2 a_1 -1 + k_0}. \ee
By comparing \eq{poleq} with \eq{acsol}, we can find the relation
between the vev for current operator and the background electric
field \be \bra {J_x} \ket = \frac{c_2}{i \omega c_1} \frac{\tau_{\rm eff} (2 \pi \a')^2}{(2 a_1
-1)^{\frac{2 a_1 -1 +k_0}{2 a_1 -1}} a_0^{\frac{2 (2 a_1 -1 +k_0)}{2
a_1 -1}}} E_x . \ee Therefore, the AC electric conductivity becomes
\bea \s_{AC} \equiv \frac{\bra {J_x} \ket}{E_x} = \frac{c_2}{i\omega
c_1} \frac{\tau_{\rm eff} (2 \pi \a')^2}{(2 a_1 -1)^{\frac{2 a_1 -1 +k_0}{2 a_1 -1}}
a_0^{\frac{2 (2 a_1 -1 +k_0)}{2 a_1 -1}}}  , \eea where the
numerical value of $\frac{c_2}{c_1}$ is given by solving the
Sch\"{o}dinger equation numerically with initial conditions
determined from the horizon solution \eq{horsol}, which satisfies
the incoming boundary condition. In Fig. \ref{fig1}, we shows the
real and imaginary conductivity depending on the charge density $C
\sim J^t$. As shown in figure, at given $\omega$ the real and
imaginary conductivity increases and decreases respectively, as the
charge density increases.

Interestingly, even for zero density in Fig. \ref{fig1} there exists non-zero conductivity, which
may be related to the effect of the pair creation of charged particles. Usually,
as the energy goes up more charged particles can be created which can explain increasing of the
real conductivity at the high energy region. Another interesting point is that at the high density
and low frequency regime (see the case for $C=4$) the conductivity decreases as the frequency
increases. This aspect would be explained as the follow: in this regime the pair creation
of the charged particles generates induced electric field which diminishes
the background electric field. Since the charged carrier moves slowly due to the weakened
electric field, the amount of the charged carrier current also grows smaller. This
can explain the drop of the real conductivity in the high density and small energy regime.
In the large energy case above the some critical frequency,
the effect of the pair creation of charged particles would be more dominant, so
it makes the real conductivity increases as the energy increases.

\section{Massive charge carriers}
We will consider the effects of massive charge carriers in this
section, which are more closely related to model building for real
-world strange metals. In this case, the energy gap is large
compared to the temperature: $E_{\rm gap}\gg T$. When translated
into the language of probe D-branes, the massive charge carriers
correspond to flavor branes wrapping internal cycles, whose volumes
vary with radial direction~\cite{Karch:2002sh, Kobayashi:2006sb}.

As was pointed out in~\cite{Karch:2002sh, Kobayashi:2006sb}, at
finite temperature the flavor brane shrinks to a point at $v=v_{0}$
for large enough mass. In this case the charge carriers correspond
to strings stretching from the flavor brane from $v=v_{0}$ to the
black hole horizon $v=v_{+}$. In a dilute limit one can consider a
small density of such strings and ignore the backreaction. For
larger densities the backreaction cannot be neglected and the
resulting configuration is that the brane forms a ``spike'' in place
of the strings. We will study the dilute limit first and then take
the backreaction into account.

\subsection{Conductivity in the dilute regime}
The massive charge carriers can be treated as strings stretching
from the probe brane to the black hole horizon in the dilute limit,
while there are no interactions between the strings. In this limit
the probe brane is described by the zero-density solution, which is
has a cigar shape with $v=v_{0}$ at the tip.

Let us take the static gauge $t=\tau, v=\sigma$. When expanded to
quadratic order for the transverse fluctuations, the Nambu-Goto
action
\begin{equation}
S_{\rm NG}=-T\int d\tau d\sigma\sqrt{-{\rm det}h_{ab}}+\int A
\end{equation}
gives the following field equation
\begin{equation}
-\frac{1}{\alpha^{\prime}}\frac{a^{2}_{0}f(v)}{v^{2a_{1}+2b_{1}-2}}
\partial_{v}x^{i}+F_{i0}+F_{ij}\dot{x}^{j}=0.
\end{equation}
Notice that we have incorporated the surface terms.

In the zero-frequency limit, the field equation can be easily
integrated out
\begin{equation}
x^{i}=V^{i}(t+ \frac{1}{v_+^{2 b_1}} \int^{v}\frac{u^{2a_{1}+2b_{1}-2}}{a_{0}^{2}f(u)}du),
\end{equation}
where $V^{i}$ is an integration constant and the relative
normalization of the two terms is fixed by imposing incoming boundary
conditions at the horizon. By assuming  $v_{0}\ll v_{+}$, at the
boundary we can obtain
\begin{equation}
\label{4eq4} \frac{1}{v_+^{2 b_1}}
\frac{1}{\alpha^{\prime}}V^{i}=F_{i0}+F_{ij}V^{j},
\end{equation}
According to Drude's law we have the following relation
\begin{equation}
\frac{m}{\tau}\propto\frac{1}{\alpha^{\prime}}T^{\frac{2 b_1}{2 a_1 -1}},
\end{equation}
Therefore the DC conductivity in the dilute limit of the massive
charge carriers is given by
\begin{equation}
\sigma=\frac{\tau}{m}J^{t}\propto\frac{J^{t}}{T^{\frac{2 b_1}{2 a_1 -1}}}.
\end{equation}

Next we consider the AC case. Before performing the calculations we
shall determine one scale of interest, the energy gap, which
corresponds to the mass of the string stretching from $v_{0}$ to
$v_{+}$,
\begin{equation}
E_{\rm
gap}=\int^{v_{+}}_{v_{0}}dv\sqrt{-g_{tt}g_{vv}}/\alpha^{\prime}=
\frac{L^{2}}{\alpha^{\prime}v_{0}} ,
\end{equation}
where another natural scale $\omega_{0}=v_{0}^{-1}$ corresponding
to the energy scale of bulk excitations at $v_{0}$, is introduced.

The bulk equation of motion for $x(v,t)={\rm
Re}(X_{\omega}(v)e^{-i\omega t})$ reads
\begin{equation}
\partial_{v}(\frac{a_{0}^{2}f(v)}{v^{2a_{1}+2b_{1}-2}}\partial_{v}X_{\omega})
=-\omega^{2}\frac{v^{2a_{1}-2b_{1}-2}}{a^{2}_{0}f(v)}X_{\omega}.
\end{equation}
Now consider the case of zero magnetic field $F_{10}=E, F_{ij}=0$,
at the boundary $v=v_{0}$ the surface term gives
\begin{equation}
\frac{1}{\alpha^{\prime}}\frac{a^{2}_{0}f(v_{0})}{v^{2a_{1}+2b_{1}-2}_{0}}
\partial_{v}X_{\omega}(v_{0})=E.
\end{equation}
Then the conductivity can be evaluated as
\begin{equation}
\sigma=\frac{J^{t}V_{\omega}(v_{0})}{E}=\frac{i\omega
J^{t}X_{\omega}(v_{0})}{E}=\frac{i\omega\alpha^{\prime}J^{t}X_{\omega}
(v_{0})v^{2a_{1}+2b_{1}-2}}{a^{2}_{0}f(v_{0})\partial_{\omega}X(v_{0})},
\end{equation}

Consider the very high frequency limit $\omega\gg\omega_{0}$, in
this case the WKB approximation can be adopted in the whole range
$v_{0}\leq v\leq v_{+}$. The leading WKB solution is given by
\begin{equation}
X_{\omega}(v)\approx C_{1}e^{-i\int^{v}_{v_{0}}\frac{\omega
v^{2a_{1}-2}}{a^{2}_{0}f(v)}}+C_{2}e^{i\int^{v}_{v_{0}}\frac{\omega
v^{2a_{1}-2}}{a^{2}_{0}f(v)}}.
\end{equation}
Then the conductivity in this regime is
\begin{equation}
\partial_{v}X_{\omega}(v_{0})\sim\frac{i\omega
v^{2a_{1}-2}_{0}}{a_{0}^{2}f(v_{0})}X_{\omega}(v_{0}),
\end{equation}
\begin{equation}
\sigma_{\rm WKB}=\alpha^{\prime}L^{-2}J^{t}v^{2b_{1}-2}_{0}.
\end{equation}

Next we consider the limit $T\ll\omega\ll\omega_{0}$, For $v\ll
v_{+}$, we have $f\approx1$ and the equation of motion turns out to
be
\begin{equation}
\partial^{2}_{v}X_{\omega}-(2a_{1}+2b_{1}-2)\frac{1}{v}\partial_{v}X_{\omega}+
\frac{\omega^{2}}{a^{4}_{0}}v^{4a_{1}-4}X_{\omega}=0.
\end{equation}
The solution is given in terms of the Bessel function
\begin{equation}
X_{\omega}(v)=v^{\xi\kappa}\omega^{\xi}H^{(1)}_{\xi}(\frac{\omega
v^{\kappa}}{\kappa a^{2}_{0}}),
\end{equation}
where \begin{equation} \label{4eq16}
\kappa=a_{1}-b_{1},~~~\xi=\frac{2a_{1}+2b_{1}-1}{2a_{1}-2b_{1}}.
\end{equation}
In the regime $\omega\ll E_{\rm gap}$, we can expand the solution
for small argument
\begin{equation}
X_{\omega}\propto1-\frac{1}{\Gamma(-\xi)}(\frac{\omega
v^{\kappa}}{2\kappa a^{2}_{0}})^{2}+e^{i\pi\xi}(\frac{\omega
v^{\kappa}}{2\kappa a^{2}_{0}})^{2\xi}.
\end{equation}
The following situations should be discussed separately when
evaluating the conductivity. The conductivity is given by the
following formula when the second term dominates
\begin{equation}
\sigma=-i\omega^{-1}2\kappa\Gamma(-\xi)\alpha^{\prime}L^{-2}J^{t}a^{2}_{0}v^{4b_{1}-1}_{0},
\end{equation}
which is a Drude result. When the third term dominates, the
conductivity possesses a nontrivial scaling
\begin{equation}
\sigma=\frac{a^{4\xi-2}_{0}}{2\xi}(2\kappa)^{2\xi}\alpha^{\prime}L^{-2}
J^{t}e^{i\pi(\frac{1}{2}-\xi)}
\omega^{1-2\xi}.
\end{equation}

In sum, we have the following scaling behaviors for the resistivity
and conductivity in the dilute regime,
\begin{equation}
\rho=\frac{1}{\sigma}\sim T^{\frac{2b_{1}}{2b_{1}-2a_{1}}},~~~
\sigma(\omega)\sim\omega^{1-2\xi},
\end{equation}
where $\xi$ is given in~(\ref{4eq16}). Notice that in real-world
strange metals,
\begin{equation}
\rho\sim T^{\nu_{1}},~~~ \sigma(\omega)\sim\omega^{-\nu_{2}},
\end{equation}
where $\nu_{1}\approx1,\nu_{2}\approx0.65$~\cite{vandeMarel:2003wn}.
Therefore if we require our dual gravitational background to exhibit
the same scaling behavior, we have to set
\begin{equation}
\alpha=\pm0.293491,~~~\eta=\pm1.45201.
\end{equation}
The consistency of the above choices for the parameters has been
verified.
\subsection{Conductivity at finite densities}
When the backreaction of the massive charge carriers cannot be
neglected, we should introduce an additional scalar field which
corresponds to the ``mass'' operator in the boundary field theory.
The volume of the internal cycle wrapped by the probe brane is
determined by this scalar field. In this case the DBI action becomes
\begin{equation}
S_{q}=-\tau_{\rm eff}\int
dtd^{2}xdvv^{-k_{0}}V(\theta)^{n}\sqrt{g_{xx}}\sqrt{-g_{tt}g_{xx}g_{\sigma\sigma}-(2\pi
\alpha^{\prime})^{2}(g_{\sigma\sigma}E^{2}+g_{xx}\Phi^{\prime2}+g_{tt}h^{\prime2})},
\end{equation}
where $V(\theta)^{n}$ denotes the volume of the $n$-dimensional
submanifold wrapped by the probe brane. Notice that here we have
introduced the worldvolume gauge field as follows
\begin{equation}
A=\Phi(v)dt+(-Et+h(v))dx
\end{equation}
and the induced metric component is given by
$g_{\sigma\sigma}=g_{vv}+\theta^{\prime2}$.

We can obtain the following constants of motion
\begin{equation}
C=\frac{-g_{xx}^{3/2}\Phi^{\prime}v^{-k_{0}}V(\theta)^{n}}{\sqrt{-g_{tt}g_{xx}g_{\sigma\sigma}-(2\pi
\alpha^{\prime})^{2}(g_{\sigma\sigma}E^{2}+g_{xx}\Phi^{\prime2}+g_{tt}h^{\prime2})}},
\end{equation}
\begin{equation}
H=\frac{-g_{tt}g_{xx}^{1/2}h^{\prime}v^{-k_{0}}V(\theta)^{n}}{\sqrt{-g_{tt}g_{xx}g_{\sigma\sigma}-(2\pi
\alpha^{\prime})^{2}(g_{\sigma\sigma}E^{2}+g_{xx}\Phi^{\prime2}+g_{tt}h^{\prime2})}}.
\end{equation}
Therefore the on-shell DBI action turns out to be
\begin{equation}
S_{q}=-\tau_{\rm eff}\int
dtd^{2}xdvv^{-2k_{0}}V(\theta)^{2n}g_{xx}^{3/2}\sqrt{-g_{tt}g_{\sigma\sigma}}GF^{1/2},
\end{equation}
where
\begin{equation}
GF=\frac{g_{tt}g_{xx}+(2\pi\alpha^{\prime})^{2}E^{2}}{(2\pi\alpha^{\prime})^{2}(C^{2}g_{tt}+H^{2}g_{xx})
+g^{2}_{xx}g_{tt}v^{-2k_{0}}V(\theta)^{2n}}.
\end{equation}
The background profile of the probe brane $\theta(v)$ is determined
by
\begin{equation}
\frac{d}{dv}[v^{-2k_{0}}V(\theta)^{2n}g^{3/2}_{xx}\sqrt{\frac{-g_{tt}}{g_{\sigma\sigma}}}
\theta^{\prime}GF^{1/2}]-\frac{d}{d\theta}[v^{-2k_{0}}V(\theta)^{2n}g^{3/2}_{xx}
\sqrt{-g_{tt}g_{\sigma\sigma}}GF^{1/2}]=0.
\end{equation}

Finally, following the same steps straightforwardly, we obtain the
DC conductivity
\begin{equation}
\sigma=\sqrt{(2\pi\alpha^{\prime})^{4}\tau_{\rm
eff}^{2}v^{-2k_{0}}_{\ast}V(\theta_{\ast})^{2n}+(\frac{2\pi\alpha^{\prime}}{L^{2}})
^{2}(J^{t})^{2}v^{4b_{1}}_{\ast}}.
\end{equation}
Note that in the massless limit $V(\theta)~\rightarrow~1$, the above
result reduces to the one obtained in previous
section~(\ref{3eq18}).

The calculations of the AC conductivity are similar to the massless
case, that is, we will study the Maxwell fluctuations up to
quadratic order and consider zero momentum $k=0$. One crucial
difference is that here we have non-trivial background profile
$\theta(v)$. The linearized equation for the longitudinal
fluctuation is given by
\begin{equation}
\label{4eq28}
\partial_{v}(v^{-k_{0}}V(\theta)^{n}\gamma\sqrt{\frac{-g_{tt}}{g_{\sigma\sigma}}}A_{x}^{\prime})
=-v^{-k_{0}}V(\theta)^{n}\gamma\sqrt{\frac{g_{\sigma\sigma}}{-g_{tt}}}\omega^{2}A_{x},
\end{equation}
where
\begin{equation}
\gamma\equiv\sqrt{1
+(2\pi\alpha^{\prime})^{2}\frac{C^{2}v^{2k_{0}+4b_{1}}}{V(\theta)^{2n}}}.
\end{equation}

The equation for the fluctuation can still be transformed into a
Schr\"{o}dinger form. After defining
\begin{equation}
A_{x}=(v^{-k_{0}}V(\theta)^{n}\gamma)^{-1/2}\Psi,
~~~\frac{d}{dv}=\sqrt{\frac{g_{\sigma\sigma}}{-g_{tt}}}\frac{d}{ds},
\end{equation}
(\ref{4eq28}) becomes
\begin{equation}
-\frac{d^{2}}{ds^{2}}\Psi+U\Psi=\omega^{2}\Psi,
\end{equation}
with the following effective potential
\begin{equation}
U=\frac{1}{2}\frac{1}{\sqrt{v^{-k_{0}}V(\theta)^{n}\gamma}}\frac{d}{ds}[\frac{1}{\sqrt
{v^{-k_{0}}V(\theta)^{n}\gamma}}\frac{d}{ds}(v^{-k_{0}}V(\theta)^{n}\gamma)].
\end{equation}

At first sight one may consider that we can perform similar
calculations as for the massless case. However, one key point in the
calculations of~\cite{Hartnoll:2009ns} was that in a wide
range $v_{0}<v<v_{+}$, the profile of the embedding was
approximately constant. This behavior was confirmed by numerical
calculations and played a crucial role in simplifying the corresponding
Schr\"{o}dinger equation. Unfortunately, here we cannot find such a
constant behavior for the profile of the embedding. The profile
turns out to be singular in the near horizon region. Thus we cannot
simplify the Schr\"{o}dinger equation considerably to find the
analytic solutions. It might be related to the non-trivial dilaton
field in the DBI action and we expect that such singular behavior
may be cured in realistic D-brane configurations, which might enable
us to calculate the transport coefficients.

\section{Summary and discussion}
The holographic model building of ``strange'' metals was initiated
in~\cite{Hartnoll:2009ns}, where the authors considered Lifshitz
black holes as the background and probe D-branes as charge carriers.
In this paper we generalize the analysis to backgrounds with
anisotropic scaling symmetry, which are solutions of the
Einstein-Maxwell-dilaton theory coupled with a Liouville potential.
For massless charge carriers, we obtain the DC conductivity and DC
Hall conductivity by applying the approach proposed
in~\cite{Karch:2007pd} and~\cite{O'Bannon:2007in}. The results can
reproduce those obtained in~\cite{Hartnoll:2009ns} in certain
specific limits. We also calculate the AC conductivity by
transforming the corresponding equation of motion into the
Schr\"{o}dinger equation. For massive charge carriers, the DC and AC
conductivities are also obtained in the dilute limit. When the dual
gravity background exhibits the scaling behaviors for strange
metals, the parameters in the action should take the following
values, $\alpha=\pm0.293491$  and $\eta=\pm1.45201$. We also obtain the
DC conductivity at finite density. One thing we are not able to
analyze is the AC conductivity at finite density, which may be due
to the singular behavior of the profile of the brane embedding. We
expect that such a difficulty may be cured in realistic D-brane
configurations.

In the final stage of this work we noticed that some overlaps
appeared in~\cite{Charmousis:2010zz}. They clarified the landscape
of effective holographic theories in view of condensed matter
applications by studying the thermodynamics, spectra and
conductivities of several classes of charged dilatonic black hole
solutions that include the charge density backreaction fully. In
particular, they obtained the DC and AC conductivities in the
context of the DBI action. It should be pointed out that one crucial
difference is that we consider the solution as the global solution,
while they imposed some constraints to ensure that the solutions
were asymptotically AdS. Another topic which was not included in
their paper is the DC Hall conductivity. However, we believe that
our results are consistent with theirs.

One important further direction is to study the fermionic nature of
the dual gravity background, as Fermi surfaces and other related
observables are crucial for the description of strange metals. One
may perform the analysis along the line of~\cite{Lee:2008xf,
Liu:2009dm, Cubrovic:2009ye, Faulkner:2009wj}. One point is that in
those papers the asymptotic geometry was AdS$_{d+2}$ and the near
horizon geometry contained an AdS$_2$ part, which played a central
role in the investigations. Although the asymptotic and/or near
horizon geometries possess anisotropic scaling symmetry, it is
believed that similar analysis can be performed by making use of the
``matching'' technique adopted in the above mentioned references.
Furthermore, one may examine the fermionic correlators in the
background studied in this paper, following~\cite{Gubser:2009dt,
Faulkner:2009am, Faulkner:2010da}. We expect to study such
interesting problems in the future.

\bigskip \goodbreak \centerline{\bf Acknowledgements}
\noindent DWP would like to thank Andy O'Bannon for helpful
discussions. This work was supported by the National Research
Foundation of Korea(NRF) grant funded by the Korea government(MEST)
through the Center for Quantum Spacetime(CQUeST) of Sogang
University with grant number 2005-0049409.




\begin{thebibliography}{99}
\addcontentsline{toc}{section}{References}


\bibitem{Maldacena:1997re}
  J.~M.~Maldacena,
  ``The large N limit of superconformal field theories and supergravity,''
  Adv.\ Theor.\ Math.\ Phys.\  {\bf 2}, 231 (1998)
  [Int.\ J.\ Theor.\ Phys.\  {\bf 38}, 1113 (1999)]
  [arXiv:hep-th/9711200].\\
  S.~S.~Gubser, I.~R.~Klebanov and A.~M.~Polyakov,
  ``Gauge theory correlators from non-critical string theory,''
  Phys.\ Lett.\  B {\bf 428}, 105 (1998)
  [arXiv:hep-th/9802109].\\
  E.~Witten,
  ``Anti-de Sitter space and holography,''
  Adv.\ Theor.\ Math.\ Phys.\  {\bf 2}, 253 (1998)
  [arXiv:hep-th/9802150].

\bibitem{Aharony:1999ti}
  O.~Aharony, S.~S.~Gubser, J.~M.~Maldacena, H.~Ooguri and Y.~Oz,
  ``Large N field theories, string theory and gravity,''
  Phys.\ Rept.\  {\bf 323}, 183 (2000)
  [arXiv:hep-th/9905111].

\bibitem{Hartnoll:2009sz}
  S.~A.~Hartnoll,
  ``Lectures on holographic methods for condensed matter physics,''
  Class.\ Quant.\ Grav.\  {\bf 26}, 224002 (2009)
  [arXiv:0903.3246 [hep-th]].\\
  C.~P.~Herzog,
  ``Lectures on Holographic Superfluidity and Superconductivity,''
  J.\ Phys.\ A  {\bf 42}, 343001 (2009)
  [arXiv:0904.1975 [hep-th]].\\
   J.~McGreevy,
  ``Holographic duality with a view toward many-body physics,''
  arXiv:0909.0518 [hep-th].\\
  G.~T.~Horowitz,
  ``Introduction to Holographic Superconductors,''
  arXiv:1002.1722 [hep-th].\\
  S.~Sachdev,
  ``Condensed matter and AdS/CFT,''
  arXiv:1002.2947 [hep-th].
\bibitem{Gubser:2008px}
  S.~S.~Gubser,
  ``Breaking an Abelian gauge symmetry near a black hole horizon,''
  Phys.\ Rev.\  D {\bf 78}, 065034 (2008)
  [arXiv:0801.2977 [hep-th]].
\bibitem{Hartnoll:2008vx}
  S.~A.~Hartnoll, C.~P.~Herzog and G.~T.~Horowitz,
  ``Building a Holographic Superconductor,''
  Phys.\ Rev.\ Lett.\  {\bf 101}, 031601 (2008)
  [arXiv:0803.3295 [hep-th]].
\bibitem{Gubser:2008wv}
  S.~S.~Gubser and S.~S.~Pufu,
  ``The gravity dual of a p-wave superconductor,''
  JHEP {\bf 0811}, 033 (2008)
  [arXiv:0805.2960 [hep-th]].
\bibitem{Hartnoll:2008kx}
  S.~A.~Hartnoll, C.~P.~Herzog and G.~T.~Horowitz,
  ``Holographic Superconductors,''
  JHEP {\bf 0812}, 015 (2008)
  [arXiv:0810.1563 [hep-th]].
\bibitem{Chen:2010mk}
  J.~W.~Chen, Y.~J.~Kao, D.~Maity, W.~Y.~Wen and C.~P.~Yeh,
  ``Towards A Holographic Model of D-Wave Superconductors,''
  Phys.\ Rev.\  D {\bf 81}, 106008 (2010)
  [arXiv:1003.2991 [hep-th]].
\bibitem{Herzog:2010vz}
  C.~P.~Herzog,
  ``An Analytic Holographic Superconductor,''
  arXiv:1003.3278 [hep-th].

\bibitem{Ammon:2008fc}
  M.~Ammon, J.~Erdmenger, M.~Kaminski and P.~Kerner,
  ``Superconductivity from gauge/gravity duality with flavor,''
  Phys.\ Lett.\  B {\bf 680}, 516 (2009)
  [arXiv:0810.2316 [hep-th]].
\bibitem{Ammon:2009fe}
  M.~Ammon, J.~Erdmenger, M.~Kaminski and P.~Kerner,
  ``Flavor Superconductivity from Gauge/Gravity Duality,''
  JHEP {\bf 0910}, 067 (2009)
  [arXiv:0903.1864 [hep-th]].
\bibitem{Gubser:2009qm}
  S.~S.~Gubser, C.~P.~Herzog, S.~S.~Pufu and T.~Tesileanu,
  ``Superconductors from Superstrings,''
  Phys.\ Rev.\ Lett.\  {\bf 103}, 141601 (2009)
  [arXiv:0907.3510 [hep-th]].
\bibitem{Gauntlett:2009dn}
  J.~P.~Gauntlett, J.~Sonner and T.~Wiseman,
  ``Holographic superconductivity in M-Theory,''
  Phys.\ Rev.\ Lett.\  {\bf 103}, 151601 (2009)
  [arXiv:0907.3796 [hep-th]].
\bibitem{Kaminski:2010zu}
  M.~Kaminski,
  ``Flavor Superconductivity and Superfluidity,''
  arXiv:1002.4886 [hep-th].
\bibitem{Hartnoll:2009ns}
  S.~A.~Hartnoll, J.~Polchinski, E.~Silverstein and D.~Tong,
  ``Towards strange metallic holography,''
  JHEP {\bf 1004}, 120 (2010)
  [arXiv:0912.1061 [hep-th]].

\bibitem{Gibbons:1987ps}
  G.~W.~Gibbons and K.~i.~Maeda,
  ``Black Holes And Membranes In Higher Dimensional Theories With Dilaton
  Fields,''
  Nucl.\ Phys.\  B {\bf 298}, 741 (1988).

\bibitem{Preskill:1991tb}
  J.~Preskill, P.~Schwarz, A.~D.~Shapere, S.~Trivedi and F.~Wilczek,
  ``Limitations on the statistical description of black holes,''
  Mod.\ Phys.\ Lett.\  A {\bf 6}, 2353 (1991).

\bibitem{Garfinkle:1990qj}
  D.~Garfinkle, G.~T.~Horowitz and A.~Strominger,
  ``Charged black holes in string theory,''
  Phys.\ Rev.\  D {\bf 43}, 3140 (1991)
  [Erratum-ibid.\  D {\bf 45}, 3888 (1992)].

\bibitem{Holzhey:1991bx}
  C.~F.~E.~Holzhey and F.~Wilczek,
  ``Black holes as elementary particles,''
  Nucl.\ Phys.\  B {\bf 380}, 447 (1992)
  [arXiv:hep-th/9202014].

\bibitem{Goldstein:2009cv}
  K.~Goldstein, S.~Kachru, S.~Prakash and S.~P.~Trivedi,
  ``Holography of Charged Dilaton Black Holes,''
  arXiv:0911.3586 [hep-th].
\bibitem{Kachru:2008yh}
  S.~Kachru, X.~Liu and M.~Mulligan,
  ``Gravity Duals of Lifshitz-like Fixed Points,''
  Phys.\ Rev.\  D {\bf 78}, 106005 (2008)
  [arXiv:0808.1725 [hep-th]].
\bibitem{Taylor:2008tg}
  M.~Taylor,
  ``Non-relativistic holography,''
  arXiv:0812.0530 [hep-th].

\bibitem{Gubser:2009qt}
  S.~S.~Gubser and F.~D.~Rocha,
  ``Peculiar properties of a charged dilatonic black hole in $AdS_{5}$,''
  Phys.\ Rev.\  D {\bf 81}, 046001 (2010)
  [arXiv:0911.2898 [hep-th]].

\bibitem{Gauntlett:2009bh}
  J.~Gauntlett, J.~Sonner and T.~Wiseman,
  ``Quantum Criticality and Holographic Superconductors in M-theory,''
  JHEP {\bf 1002}, 060 (2010)
  [arXiv:0912.0512 [hep-th]].

\bibitem{Cadoni:2009xm}
  M.~Cadoni, G.~D'Appollonio and P.~Pani,
  ``Phase transitions between Reissner-Nordstrom and dilatonic black holes in
  4D AdS spacetime,''
  arXiv:0912.3520 [hep-th].
\bibitem{Chen:2010kn}
  C.~M.~Chen and D.~W.~Pang,
  ``Holography of Charged Dilaton Black Holes in General Dimensions,''
  arXiv:1003.5064 [hep-th].
\bibitem{Charmousis:2010zz}
  C.~Charmousis, B.~Gouteraux, B.~S.~Kim, E.~Kiritsis and R.~Meyer,
  ``Effective Holographic Theories for low-temperature condensed matter
  systems,''
  arXiv:1005.4690 [hep-th].

\bibitem{Cai:1996eg}
  R.~G.~Cai and Y.~Z.~Zhang,
  ``Black plane solutions in four-dimensional spacetimes,''
  Phys.\ Rev.\  D {\bf 54}, 4891 (1996)
  [arXiv:gr-qc/9609065].\\
  R.~G.~Cai, J.~Y.~Ji and K.~S.~Soh,
  ``Topological dilaton black holes,''
  Phys.\ Rev.\  D {\bf 57}, 6547 (1998)
  [arXiv:gr-qc/9708063].\\
  C.~Charmousis, B.~Gouteraux and J.~Soda,
  ``Einstein-Maxwell-Dilaton theories with a Liouville potential,''
  Phys.\ Rev.\  D {\bf 80}, 024028 (2009)
  [arXiv:0905.3337 [gr-qc]].
\bibitem{Lee:2010xx}
B.~H.~Lee, S.~Nam, D.~W.~Pang and C.~Park, ``Conductivity in the
anisotropic background,'' arXiv: 1006.0779 [hep-th].

\bibitem{Karch:2002sh}
  A.~Karch and E.~Katz,
  ``Adding flavor to AdS/CFT,''
  JHEP {\bf 0206}, 043 (2002)
  [arXiv:hep-th/0205236].

\bibitem{Karch:2007pd}
  A.~Karch and A.~O'Bannon,
  ``Metallic AdS/CFT,''
  JHEP {\bf 0709}, 024 (2007)
  [arXiv:0705.3870 [hep-th]].

\bibitem{O'Bannon:2007in}
  A.~O'Bannon,
  ``Hall Conductivity of Flavor Fields from AdS/CFT,''
  Phys.\ Rev.\  D {\bf 76}, 086007 (2007)
  [arXiv:0708.1994 [hep-th]].
\bibitem{Kobayashi:2006sb}
  S.~Kobayashi, D.~Mateos, S.~Matsuura, R.~C.~Myers and R.~M.~Thomson,
  ``Holographic phase transitions at finite baryon density,''
  JHEP {\bf 0702}, 016 (2007)
  [arXiv:hep-th/0611099].

\bibitem{Gubser:2008wz}
  S.~S.~Gubser and F.~D.~Rocha,
  ``The gravity dual to a quantum critical point with spontaneous symmetry
  breaking,''
  Phys.\ Rev.\ Lett.\  {\bf 102}, 061601 (2009)
  [arXiv:0807.1737 [hep-th]].
\bibitem{Horowitz:2009ij}
  G.~T.~Horowitz and M.~M.~Roberts,
  ``Zero Temperature Limit of Holographic Superconductors,''
  JHEP {\bf 0911}, 015 (2009)
  [arXiv:0908.3677 [hep-th]].

\bibitem{vandeMarel:2003wn}
  D.~van de Marel {\it et al.},
  ``Quantum critical behaviour in a high-tc superconductor,''
  Nature {\bf 425} (2003) 271.


\bibitem{Lee:2008xf}
  S.~S.~Lee,
  ``A Non-Fermi Liquid from a Charged Black Hole: A Critical Fermi Ball,''
  Phys.\ Rev.\  D {\bf 79}, 086006 (2009)
  [arXiv:0809.3402 [hep-th]].

\bibitem{Liu:2009dm}
  H.~Liu, J.~McGreevy and D.~Vegh,
  ``Non-Fermi liquids from holography,''
  arXiv:0903.2477 [hep-th].

\bibitem{Cubrovic:2009ye}
  M.~Cubrovic, J.~Zaanen and K.~Schalm,
  ``String Theory, Quantum Phase Transitions and the Emergent Fermi-Liquid,''
  Science {\bf 325}, 439 (2009)
  [arXiv:0904.1993 [hep-th]].

\bibitem{Faulkner:2009wj}
  T.~Faulkner, H.~Liu, J.~McGreevy and D.~Vegh,
  ``Emergent quantum criticality, Fermi surfaces, and AdS2,''
  arXiv:0907.2694 [hep-th].

\bibitem{Gubser:2009dt}
  S.~S.~Gubser, F.~D.~Rocha and P.~Talavera,
  ``Normalizable fermion modes in a holographic superconductor,''
  arXiv:0911.3632 [hep-th].
\bibitem{Faulkner:2009am}
  T.~Faulkner, G.~T.~Horowitz, J.~McGreevy, M.~M.~Roberts and D.~Vegh,
  ``Photoemission 'experiments' on holographic superconductors,''
  JHEP {\bf 1003}, 121 (2010)
  [arXiv:0911.3402 [hep-th]].
\bibitem{Faulkner:2010da}
  T.~Faulkner, N.~Iqbal, H.~Liu, J.~McGreevy and D.~Vegh,
  ``From black holes to strange metals,''
  arXiv:1003.1728 [hep-th].




\end{thebibliography}
\end{document}